\documentclass[12pt]{article}

\usepackage{amsmath}
\usepackage{amssymb}
\usepackage{amsfonts}
\usepackage{amscd}
\usepackage{amsbsy}
\usepackage{amsthm}
\usepackage{latexsym}
\usepackage{graphicx}
\usepackage{slashed}
\usepackage{color}

\usepackage[utf8]{inputenc}
\usepackage[T1]{fontenc}
\usepackage{mathrsfs}
\usepackage{subfigure}

\usepackage[hyperindex]{hyperref}
\hypersetup{colorlinks=true,linkcolor=blue,citecolor=blue,urlcolor=blue}

\renewcommand{\baselinestretch}{1.2}
\def\beq{\begin{eqnarray}}
\def\eeq{\end{eqnarray}}
\newcommand{\nn}{\nonumber}

\def\ln{\,\mbox{ln}\,}

\def\al{\alpha}
\def\be{\beta}

\def\la{\lambda}
\def\na{\nabla}
\def\pa{\partial}

\def\si{\sigma}

\def\Ga{\Gamma}

\def\La{\Lambda}

\begin{document}

\begin{center}

{\large\bf Scale-dependent cosmology from effective quantum gravity
\\
in the invariant framework}

\vskip 6mm

\textbf{Nicolas R. Bertini$^{1}$},
\textbf{Davi C. Rodrigues$^{2, 3, 4}$}
\ and \
\textbf{Ilya L. Shapiro$^{1}$}

\vskip 6mm

{\sl $^{1}$ Departamento de F\'{\i}sica, ICE,
Universidade Federal de Juiz de Fora \\ Campus Universit\'{a}rio - Juiz de Fora, 36036-900, MG, Brazil}\\
{\sl $^{2}$ Departamento de Física \& Cosmo-Ufes, Universidade Federal do Espírito Santo, 29075-910, Vitória, ES, Brazil}\\
{\sl $^{3}$ Institut für Theoretische Physik, Universität Heidelberg, \\ Philosophenweg 16, 69120. Heidelberg, Germany}\\
{\sl $^{4}$ Instituto de Física, Universidade Federal do Rio de Janeiro, \\21941-972, Rio de Janeiro, RJ, Brazil}\\

\end{center}
\vskip 6mm

\centerline{\textbf{Abstract}}
\vskip 1mm

\begin{quotation}
\noindent
We explore the possibility of a consistent cosmology based on the
gauge-fixing independent running of the gravitational and cosmological
constants ($G$ and $\Lambda$) in the framework of effective quantum gravity. 
In particular, their running in this framework was found to satisfy $G \propto \Lambda^4$.
In the
cosmological setting, the covariance of the theory provides energy
conservation relations, which are impossible to satisfy with the
unique scale parameter. However, by introducing the second
sub-dominant scale corresponding to the higher-loop corrections and
higher-derivative terms, one can close the system of equations for
the running of parameters and arrive at the consistent cosmological
solutions. 
This approach yields a change in the cosmological expansion history that affects the ratio of the Hubble parameter today to the Hubble parameter at high redshift.
\vskip 3mm

\vskip 2mm


\end{quotation}

\section{Introduction}
\label{SecIntro}

The idea to apply the renormalization group running to the
gravitational physics has a long history. One can mention, e.g., the
works in this direction developed in
\cite{antmot,Bertol,Eli-94,CC-nova,Reuter:2007de}, based on
very different approaches and
frameworks.
The reason is that it cannot be ruled
out that the equation of state for the Dark Energy may
be observationally confirmed different from $-1$ . From
the quantum field theory (QFT) viewpoint, the main candidate
for the Dark Energy is the cosmological constant (CC), and in this
case, the deviation from the canonical $-1$ can be a consequence
of the IR running of the CC
or the Newton constant, or both. Taking this into account, it is
important to explore the theoretical possibility of such a running
in different possible QFT or quantum gravity frameworks,

Depending on the theoretical background of renormalization group
running, the applications to cosmology can be, in particular,
separated into the ones based on semiclassical 
\cite{Babic2002,CC-nova,DCCrun} or the proper effects of quantized 
gravity. Most of the existing models of running vacuum cosmology
are based on the semiclassical version (see, e.g., 
\cite{CC-astron-2003,CC-Gruni,Basilakos2019,Jhonny-2020,BWD}
and references therein).
In the present work, we apply the perturbative quantum 
gravity running to cosmology and, as we shall see in what follows, 
this leads to specific kind of theoretical problem.
The first applications of such a running \cite{antmot,Bertol}, 
were based on the renormalizable quantum gravity \cite{Stelle77} 
and the corresponding running \cite{frts82,avbar86}. However, 
the renormalizability of these theories is owing to the  presence 
of massive ghosts with the masses of the Planck order
of magnitude \cite{OUP}. Even assuming that there can be a
satisfactory solution to the problem of ghosts, there remains another
obstacle for the cosmological applications. The massive degrees of
freedom contributing to the  running, should decouple in the IR, 
as it was discussed recently in \cite{AM-effect} and \cite{DGP}. 
This means, the running in the renormalizable models of quantum
gravity can be applied only at the energy scales much higher than
the typical masses, i.e., above the Planck scale. This is not 
appropriate for the applications to cosmology, thus it is 
necessary to look for an effective theory of quantum gravity.

In the effective approach, the general relativity (GR) is 
considered a universal theory of effective IR quantum gravity 
\cite{don}.\footnote{Let us note that the recent results of 
\cite{DGP} contradict this statement, but we assume that 
this standard treatment of effective quantum gravity is 
correct.}
Therefore,  the effective approach  
\cite{Burgess,Handbook-effQG} should use quantum GR 
as a basis for the cosmologically relevant running. However, 
in the corresponding quantum corrections one meets strong
gauge-fixing dependence of the beta functions. The unique 
known way to overcome this difficulty is to use the special 
version of quantum field theory, called Vilkovisky-DeWitt unique
effective action \cite{Vil-unicEA,DeWitt-ea}. The necessary
calculations were performed by Taylor and Veneziano 
\cite{TV90} and recently extended in \cite{UEA-RG,UEA-SM} 
to include the beta functions for higher derivative terms. 

Different from the flat-space QFT, in curved spacetime, one needs
a special effort to link the results obtained within the minimal 
subtraction scheme of renormalization with a given physical
situation. The application of the IR running for the cosmological
(and other gravitational) problems requires identification of the
artificial renormalization group scale $\mu$ with a certain physical
quantity. Such an identifications are not a trivial issue, which can
be resolved in different ways 
\cite{Shapiro-Sola-1999,CC-nova,Babic2002}. 
From the QFT viewpoint, $\mu$ should be associates
with a dimensional parameter characterizing the energy of the
gravitational field. For instance, in the description of expanding 
Universe it can be the Hubble parameter $H$, or its time derivative 
$\dot{H}$.
In this case,  the QFT-based arguments
provide the result which is equivalent to the specially designed
scale-setting procedure in cosmology \cite{Babic2005} and in
astrophysics \cite{RotCurve,StefDom} (see also \cite{BWD}
for the covariant identifications and further references).

A pertinent question is whether there should be only one typical
scale, or several scales. The multi-scale option has fair justification
from the phenomenological side. For instance, in cosmology one can
consider (and should do so, in principle) different energy scales for
the conformal factor of the metric and for the cosmic perturbations,
because the typical lengths in these cases may be quite different.
From the QFT and running viewpoints, this means that the energies
in external lines of the Feynman diagrams should be also different.
It is clear that the multi-scale extension may lead to some changes
in the bound on the running derived in the two types of the
cosmological models in Refs.~\cite{CCwave} and \cite{CCG}.

In the next sections, we construct the identification of scale for
the effective quantum gravity beta functions \cite{TV90,UEA-RG}.
In the semiclassical case applied to cosmology, the relation between
$\mu$-dependence
of the cosmological and Newton constants may be established by
using the conservation law. In the case of universal quantum
gravity running obtained in \cite{UEA-RG}, this is impossible
to do because both beta-functions are well-defined by the
``one-fourth'' rule. In what follows we show that this rule
becomes compatible with the conservation law only if we
include the running of the higher derivative terms with the
second, numerically suppressed scale. 

Here we consider the $G$ and $\Lambda$ running effects in the cosmological background. Such effects are assumed to be perturbations in the standard background picture, but they need not to be so small as to be comparable with the cosmological perturbations. This allows us to test the model principles within a simple and relevant physical picture: the cosmological background evolution. Aligned with this context, we use here cosmic chronometers (CsC) \cite{Jimenez:2001gg} to test the background evolution. CsC does not constitute  the most precise test, but this is a straightforward one that is quite model independent. In particular it does not depend on the cosmic distance ladder. Namely, quoting \cite{Moresco:2023zys} ``\textit{it provides a direct determination of the Hubble parameter without relying on any cosmological assumptions (apart the cosmological principal and a metric)}''. We add that, however, this approach relies on the selection of passively evolving galaxies and such selection is not trivial and may introduce systematic errors on the CsC data. Nonetheless, currently, the final result is compatible with other approaches, within the error bars. Hence, we understand this to be a simple but relevant first test for the proposed approach.  

The rest of the article is organized as follows. In the next
Sec.~\ref{sec2}, we introduce the action of our model and
construct the framework for describing the running.  The
considerations are based on the simplified form of the higher
derivative terms. Furthermore, we consider the conservation law
and the process of identifying the scale. Some technical details
are separated in the Appendix.
As an example, Sec.~\ref{sec3} describes the evolution of the
Universe dominated by dust and CC.
Finally, in  Sec.~\ref{secConc} we draw our conclusions
and discuss the further possible developments concerning the
effective quantum gravity - based running.
The notations include the signature $(-,+,+,+)$, the definition of
the Riemann tensor
$\,R_{\,.\,\be\mu\nu}^{\al}=\Ga_{\,\be\nu,\,\mu}^{\al}
-\Ga_{\,\be\mu,\,\nu}^\al \,+\, ...$\ .
The Ricci tensor is defined as $R_{\al\be}=R_{\,.\,\al\mu\be}^\mu$
and the scalar curvature as $R=R_{\,\al}^\al$.

\section{Running couplings and conservation law in cosmology}
\label{sec2}

The effective approach to quantum gravity assumes that the
Feynman rules are constructed starting with the Einstein-Hilbert
action. Then, the unique physical degree of freedom is the
graviton and there is a duality between UV and IR contributions.
The derivation of the IR effects requires only the non-local
terms in the effective action, but those can be recovered from
the UV divergences. The same concerns the beta functions,
and not only the ones for the Newton and cosmological constants, 
but also for the higher derivative terms, which can
be included into the action just to be renormalized \cite{UEA-RG}.
As we shall see in what follows, these terms are important for
closing the conservation equations, so we have to include some
expression for these terms. On the other hand, since quantum
general relativity is a non-renormalizable theory, there is an
unbounded quantity of such terms and, for practical purposes,
one has to perform a truncation. Our purpose is to explore the 
dynamics of the conformal
factor in the lowest possible approximation. This means, in the
fourth-derivative sector we can ignore the square of the Weyl
tensor, topological and surface terms, such that the general form
of the action boils down to the $R^2$. This reduction has an
additional sense because this term is also required for inflation
\cite{star,star83}. In the next order, we have many relevant
six-derivative terms. However, since these terms are relevant
only from the second-loop order, we can choose a particular
simplest version for these term.

Taking these arguments into account, we arrive at the
Einstein-Hilbert action with an additional $f(R)$ function
\beq
S[g,\Psi]
\,=\,
\frac{1}{16\pi} \int 
\Big\{\frac{1}{G}(R-2\Lambda)+f(R)\Big\} \sqrt{-g} \, d^{4}x
\,+\, S_m[g,\Psi],
\label{action}
\eeq
where $G$ and $\La$ are constants, $g$ is the determinant of the metric tensor, $f(R)$ is a polynomial of the Ricci scalar $R$ with minimum order 2 and $\Psi$ represents
any type of matter fields (independently on their nature: scalar, vector, spinor, or even particles). 
The field equations derived from the variation with respect to the metric read
\beq
G_{\alpha\beta} + G H_{\alpha\beta}
\,=\,
8\pi G(T_{\alpha\beta} - \rho_\Lambda g_{\alpha\beta}),
\label{eq:FE}
\eeq
where $G_{\alpha \beta}$ is the Einstein tensor, \ $\rho_\La = \frac{\La}{8\pi G}$
\ and
\beq
H_{\alpha\beta}
&=&
f_R R_{\alpha\beta} - \frac{1}{2}g_{\alpha\beta}f
+ (g_{\alpha\beta}\Box - \na_\al \na_\be) f_R\,,
\\
T_{\alpha\beta}
&=&
-\frac{2}{\sqrt{-g}}\frac{\delta S_m}{\delta g^{\alpha\beta}},
\label{HT}
\eeq
with $f_R = \frac{\pa f}{\pa R}$.

We consider scale-dependent effects on $G$ and $\Lambda$, such that: $G= G(\mu_1)$,  $\Lambda = \Lambda(\mu_1)$ and $f(R) = \xi_2(\mu_1) R^2 + O(R^3)$, in agreement with the results of \cite{TV90, UEA-RG}. It will be shown it is not consistent to consider a single scale $\mu_1$ in this cosmological context, hence our first step is to generalise the previous $f(R)$ expression, including a dependence on all possible scales $\mu_b$ ($b=1, 2, ...$), as follows:
\beq
f (R)\,=\,
\xi_2(\mu_1)R^{2}
+ \xi_3(\mu_b)R^{3}
+ \xi_4(\mu_b)R^4 \, .
\label{eq:f}
\eeq
The power expansion above is a simple extension of the original picture beyond $\xi_2 R^2$, it is not our purpose at this moment to present a general case. 
From the definition of $H_{\alpha \beta}$, it can be computed that
\begin{equation}
\nabla^\alpha H_{\alpha \beta} = - \frac 12 \sum_{b} 
\frac{\partial f}{\partial \mu_b} \nabla_\beta \mu_b \, .
\end{equation}
Hence, the divergence of $T^\alpha_\beta$, using Eq.~\eqref{eq:FE}, 
can be written as
\beq
\nabla_\alpha T^{\alpha}_{\;\;\beta}
&=&
\nabla_\beta \rho_\Lambda
- G^{-1}(T^{\alpha}_{\;\;\beta}
- \delta^{\alpha}_{\beta}\rho_\Lambda)\nabla_\alpha G
+ \frac{1}{8\pi G}H^{\alpha}_{\;\;\beta}\nabla_\alpha G
- \frac{1}{16\pi}\,\sum_{b}\frac{\pa f}{\pa\mu_b}\, \nabla_{\alpha}\mu_b \, .
\label{eq:div1}
\eeq
Let us additionally assume that $T_{\al\be}$ has the form of the
energy-momentum tensor of a perfect fluid, i.e.
\beq
T_{\alpha\beta} = (\rho+p)u_\alpha u_\beta + g_{\alpha\beta}\,p,
\label{fluid}
\eeq
where $p$ is the pressure, $\rho$ is the energy density, and
$u^{\al}$ is the four-velocity of the fluid. The usual conservation
law for such a fluid is
\beq
u^\be \,\na_\al T^\al_{\;\;\beta}
\,=\, - \,u^{\alpha}\nabla_{\alpha}\rho
\,-\, (\rho+p)\nabla_{\alpha}u^{\alpha} \,=\, 0.
\label{fluidcons}
\eeq
Contracting  \eqref{eq:div1} with $u^\be$, we obtain
\beq
u^{\beta}\nabla_\alpha T^{\alpha}_{\;\;\beta}
&=&
u^{\alpha}\big(\na_\al \mu_1\big)
\Big[\frac{d\rho_\Lambda}{d\mu_1}
\,+\, \frac{1}{G}\,(\rho + \rho_\Lambda)
\frac{d G}{d\mu_1}\Big]
\nn
\\
&&
+ \,\,\frac{1}{8\pi G}\,u^{\beta}(\na_\al\mu_1)\,
H^{\alpha}_{\;\;\beta}\,\,\frac{d G}{d \mu_1}
\,-\,
\frac{1}{16\pi} \,  u^\al \sum_b \big(\na_\al \mu_b\big)
\, \frac{\partial f}{\partial\mu_b}.
\label{eq:div2}
\eeq
From \eqref{eq:FE} we get
\beq
\rho+\rho_\Lambda
\,=\,
\frac{1}{8\pi G}\,
u^{\alpha}u^{\beta}
\,\big(G_{\alpha\beta}+ G H_{\alpha\beta}\big) \, .
\eeq
Thus Eq.~\eqref{eq:div2} becomes
\beq
u^{\beta}\nabla_\alpha T^{\alpha}_{\;\;\beta}
&=&
u^{\alpha} \big(\na_\al \mu_1\big)
\Big[\frac{d\rho_\La}{d\mu_1}
\,+\,
\frac{1}{8\pi G^{2}}\,\frac{d G}{d\mu_1}\,u^\si u^\la G_{\si\la}\Big]
\,-\, \frac{1}{16\pi}\, u^\al \big(\na_ \al \mu_b\big)
\,\frac{\pa f}{\pa \mu_b}
\nonumber
\\
&&
+ \,\,
\big(\na_\al \mu_1\big)
\Big[ u^{\alpha} u^{\sigma}u^{\lambda} H_{\sigma\lambda}
+ u^{\beta}H^{\alpha}_{\;\;\beta} \Big]
\,\frac{1}{8\pi G}\frac{d G}{d \mu_1}.
\label{eq:div3}
\eeq

The running of $G$, $\Lambda$ and $\xi$, was calculated in 
\cite{UEA-RG} and found to be
\beq
\Lambda(\mu_1) &=& \Lambda_0
\Big( 1+ \nu \ln \frac{\mu_1}{\overline \mu_1} \Big)^{-1/5},
\label{eq:L}
\\
G(\mu_1) &=&
G_0 \Big( 1 + \nu \ln \frac{\mu_1}{\overline \mu_1} \Big)^{-4/5},
\label{eq:G}
\\
\xi (\mu_1)
&=&
\xi_0 \,\Big( 1+ \kappa \ln \frac{\mu_1}{\overline \mu_1}\Big).
\label{eq:xi}
\eeq

In these formulas the values $\overline \mu_1$ correspond to the 
fiducial scale, and the quantities $\nu$ and $\kappa$ are constants. 
These solutions of the renormalization group equations  were 
derived in  \cite{UEA-RG} using a combination of the effective
approach to quantum gravity and the Vilkovisky-DeWitt unique
effective action.

The above relations imply that $G \propto \Lambda^4$, a relation which was not explored in previous cosmological frameworks\footnote{This relation reflects the running of the two parameters, while their absolute values have very different magnitudes.}. Indeed, for instance, in ref.~\cite{CC-Gruni} and other works based on it, $\Lambda$ depends on $\mu^2$, while $G$ depends on $\ln \mu$; hence they do not have a proportionality relation. In refs.~\cite{Bonanno:2001hi, Rodrigues:2015hba, BWD}, on the other hand, one finds $\Lambda \propto G^{-1}$, which also clearly deviates from the picture described in this framework.

We assume a homogeneous and isotropic universe, as given by the Friedmann–Lemaître–\\Robertson–Walker (FLRW) metric
\beq
ds^{2} = -dt^{2} + a^{2}(t)\delta_{ij}dx^{i}dx^{j} \, ,
\label{eq:metric}
\eeq
where $a(t)$ is the cosmological scale factor.

A well-motivated and widely used cosmological scale identification 
is with the Hubble parameter ($H = \dot a / a$) \cite{CC-nova, CC-Gruni, CCG}. 
Thus, we set
\begin{equation}
\mu_1 = H \, . \label{eq:ssH}
\end{equation}
We note that this identification is coordinate dependent. It is possible to express such quantity in coordinate-independent ways and we will comment further on this in the next section. 

In the comoving frame, the \textit{r.h.s.} of (\ref{eq:div2}) gives
non-zero result unless $f$ depends on more than one scale. Since the 
non-vanishing terms depend on $\dot H$, we set
\begin{equation}
\mu_2 = \dot{H}.\label{eq:ssdotH}
\end{equation}

On the other hand, for the full set of third and fourth-order terms
there are no quantum calculations of the running. Furthermore, there
are indications that the corresponding beta functions at higher than
the one-loop orders may be dependent on the
renormalization scheme \cite{Bern2015}. Therefore, the scale
dependencies of the corresponding parameters can be found by
using the conservation law, that means solving Eqs.~\eqref{eq:div3}
up to the first order in $\nu$. The details of this solution can be
found in Appendix~\ref{AppXi} and here we give the final results,
\beq
\xi_3(\mu_1,\mu_2)
&=&
\xi_{30}\bigg\{
1- \frac{16\pi \xi_0 \kappa}{3 \xi_{30}(2\mu_1^{2}+\mu_2)}
\ln\frac{\mu_1}{\overline \mu_1}
\nn
\\
&&
+ \,\,
\frac{\nu}{45\xi_{30}G_0 (2\mu_1^{2}+\mu_2)^{3}}
\Big[
- 2(\mu_1^{2}-\overline{\mu}_{1}^{2})
+ \Lambda_0 \ln\frac{\mu_1}{\overline \mu_1}
\Big] \bigg\},
\label{eq:xi3}
\\
\xi_4(\mu_1,\mu_2)
&=& \xi_{40}
\bigg\{ 1+ \frac{4\pi \xi_0 \kappa}{9 \xi_{40}(2\mu_1^{2}+\mu_2)^2}
\ln\frac{\mu_1}{\overline \mu_1}
\nn
\\
&&
+ \,\,
\frac{\nu}{360\xi_{40}G_0 (2\mu_1^{2}+\mu_2)^{4}}
\Big[
2(\mu_1^{2}-\overline{\mu}_{1}^{2})
- \Lambda_0 \ln\frac{\mu_1}{\overline \mu_1}  \Big] \bigg\}.
\label{eq:xi4}
\eeq
The most important qualitative detail is that without the higher
derivative terms, i.e., without $f(R)$, the solutions (\ref{eq:L})
and  (\ref{eq:G}) are incompatible with the conservation law.
From the physical side, this means if the energy exchange between
gravity and matter is forbidden (see the discussions and further
references in \cite{CC-Gruni}, \cite{Jhonny-2020} and
\cite{PoImpo}) fixing both $\La(\mu)$ and $G(\mu)$ does not
leave enough space to fit the conservation law.

\section{Application to late-time cosmology}
\label{sec3}

As a first phenomenological investigation, we consider late-time cosmology. In such scenario, the field equations are significantly simpler since the higher derivative terms in $H_{\alpha\beta}$ is suppressed with respect to the other terms in Eq.~\eqref{eq:FE}, and thus there is no explicit dependence on $\mu_2$. Adopting the metric
\eqref{eq:metric},  the 0-0 component of Eq.~\eqref{eq:FE} is
\begin{align}
3H^{2} - \Lambda(H) - \frac{3\nu}{5}\left( 2( H^{2} - \overline{\mu}_1^2) - \Lambda_0 \ln\frac{H}{\overline{\mu}_1}    \right) \,=\, 8\pi G(H) \, \rho \,,
\label{eq:friedmann}
\end{align}
where $\Lambda(H)$ and $G(H)$ are given by Eqs.~\eqref{eq:L} 
and \eqref{eq:G}, respectively. We assume that
the usual continuity equation remains satisfied, i.e.,
\begin{align}
\dot{\rho} + H(\rho+p)=0\,.
\label{eq:cont}
\end{align}

From the physical side, this means we do not allow for the energy
exchange between vacuum and matter, as this was the case in
\cite{CC-Gruni}. The reason is that such an exchange is physically
impossible at low energies \cite{OphPel} and can take place only
in the early Universe \cite{Jhonny-2020}.

Before starting the phenomenological analysis of the previous 
dynamical equations, one needs to specify the value of the reference 
scale $\overline \mu_1$. The $\overline \mu_1$ value is such that 
the renormalization group corrections become relevant only when 
$\mu_1 \not \approx \overline \mu_1$. Since such corrections are 
expected to be stronger in the primordial universe,  we take the 
future de Sitter phase  as the one without renormalization group 
corrections. Being $\overline \mu_1$ the Hubble value at the 
asymptotic de Sitter phase, we set
\begin{align}
\overline\mu_{1} = \sqrt{\frac{\Lambda_0}{3}}\,.\label{eq:mu10}
\end{align}

For a universe dominated by dust, Eq.~\eqref{eq:cont} provides 
$\rho = \rho_0 (1+z)^{3}$, where $\rho_0$ is a constant.
Therefore Eq.~\eqref{eq:friedmann} can be written as
\begin{eqnarray}
\frac{H^{2}}{H_{0}^{2}} = \Omega_\Lambda 
+ \Omega_\mathrm{m}(1+z)^{3},
\label{eq:fried}
\end{eqnarray}
where $H_0$ is the Hubble parameter at $z=0$. The expressions for 
$\Omega_\Lambda$ and $\Omega_\mathrm{m}$ are, up to the first 
order in $\nu$, given by
\begin{align}
\Omega_\Lambda = \Omega_{\Lambda 0}\left( 1- \frac{4\nu}{5} 
\ln\frac{H}{\overline\mu_1} \right), \qquad \Omega_\mathrm{m} 
= \Omega_\mathrm{m0}
\left( 1 - \frac{4\nu}{5} \ln\frac{H}{\overline\mu_1} \right),
\label{eq:omegaLandm}
\end{align}
where $\Omega_{\Lambda 0}\equiv \Lambda_0/3H_{0}^{2}$, 
$\Omega_\mathrm{m0}\equiv 8\pi G_\mathrm{N}\rho_0/3H_{0}^{2}$ and $G_\mathrm{N} \equiv G_0/(1-2\nu/5)$. With these definitions, $\overline\mu_1$ can be rewritten in the form
\begin{align}
\overline\mu_1 = H_0 \sqrt{\Omega_{\Lambda 0}}\,.\label{eq:Href}
\end{align}

For $\Lambda$CDM, one can choose to write $\Omega_{\Lambda0}$ as a function of $\Omega_{m0}$ or $\Omega_{m0}$ as a function of the latter. For the present model, only $\Omega_{m0}$ can be algebraically solved as a function of $\Omega_{\Lambda 0}$, thus we use $\Omega_{\Lambda 0}$ as the fundamental quantity. From eq.~\eqref{eq:fried} at $z=0$, one finds

\begin{equation}
\Omega_\mathrm{m0} = 1- \Omega_{\Lambda 0} - \frac{2\nu}{5}\ln{\Omega_{\Lambda 0}}   +  O(\nu^{2}) \, . \label{eq:Omegam0L0}
\end{equation}

Expanding $H$ up to the first order in the renormalization group corrections, let
\begin{equation}
  H = H^{(0)} + \nu H^{(1)}\, ,
\end{equation}
with $H(\nu = 0) = H^{(0)}$. The $H^{(1)}$ solution comes from eq.~\eqref{eq:fried} with eq.~\eqref{eq:Omegam0L0},

\begin{align}
H^{(1)}  = - \frac{H_0}{5F}\left(  F^{2} \ln \frac{F^{2}}{\Omega_{\Lambda 0}} + (1+z)^{3}\ln\Omega_{\Lambda 0} \right)
\end{align}
with
\begin{equation}
    F^2 = (1 -\Omega_{\Lambda 0}) (1+z)^3 + \Omega_{\Lambda 0} = \frac{(H^{(0)})^2}{H_0^2}\, .
\end{equation}

\begin{figure}[ht!]
\centering
\includegraphics[scale=.43]{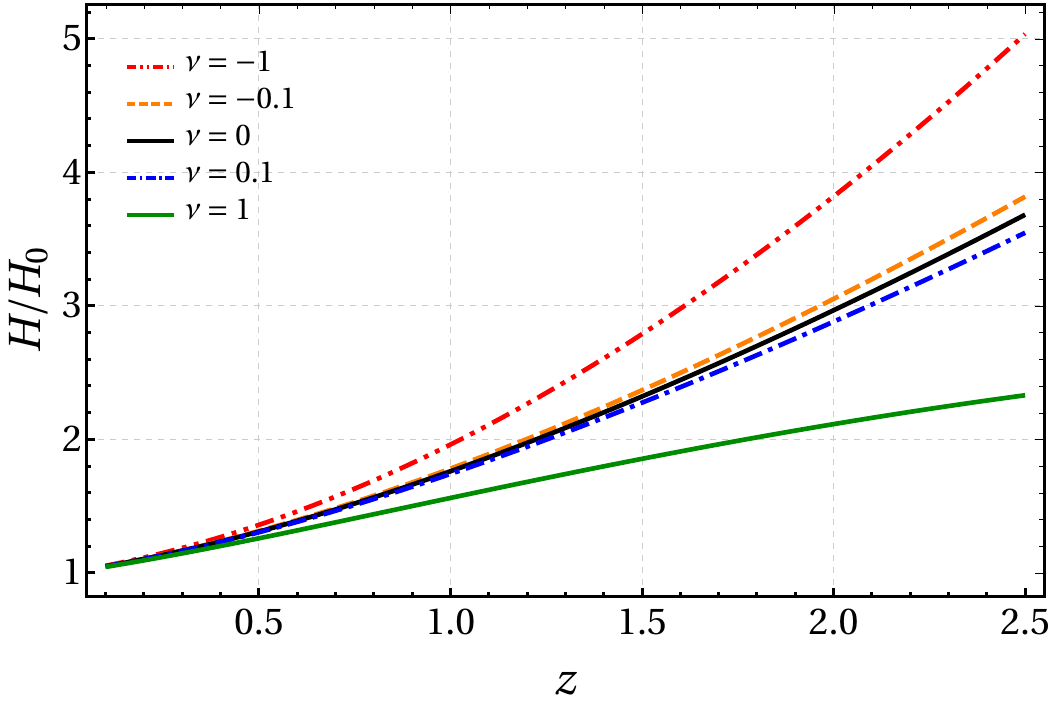}
\includegraphics[scale=.43]{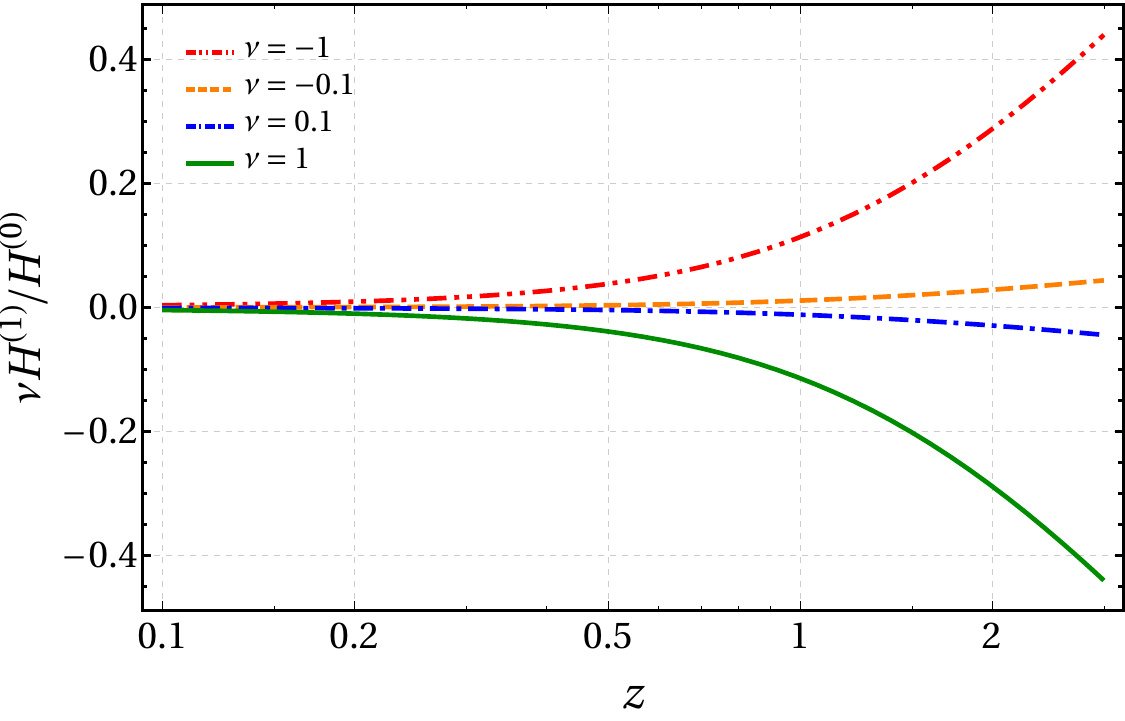}
\caption{\textbf{Left:} Hubble parameter versus redshift for different $\nu$ values. \textbf{Right:} Same as the previous plot, but for the relative deviation $\nu H^{(1)}/H^{(0)} = (H - H^{(0)})/H^{(0)}$.
}\label{figH}
\end{figure}

In Fig.~\ref{figH} we display numerical results for the Hubble  parameter for different $\nu$ values, in order to show qualitative aspects of the model. 

To assess the implications of this scenario for cosmic expansion in comparison with the standard $\Lambda$CDM model ($\nu =0$), we used the Cosmic Chronometer (CsC) data. The CsC approach \cite{Jimenez:2001gg} (see also \cite{Moresco:2023zys}) is a technique for measuring $H(z)$ which is considerably independent of the cosmological model. It is based on the measurement of $dz / dt$ and the following relation valid within FRW cosmologies,
\begin{equation}
    H(z) = - \frac{1}{1+z} \frac{dz}{dt} \, .
\end{equation}

To adjust the parameters we performed a standard $\chi^{2}$ analysis, with
\begin{align}
    \chi^{2} = \sum_{i}\frac{(H_{\rm model}(z_i) - H_{\rm obs}(z_i))^{2}}{\sigma_{i}^{2}} \, .
\end{align}
In the above,  $i$ indexes the observed redshift values ($z_i$), $H_{\rm obs}$ is observational value of the Hubble parameter,  $\sigma_i$ is the corresponding uncertainty and $H_{\rm model}$ the model Hubble parameter. The observational data that we use here are compiled in Ref.~\cite{Moresco:2023zys}.

For the present analysis, we use, at $z = 0$, $H_0 = 73.0$ km\,s$^{-1}$\,Mpc$^{-1}$. The latter value  is found from local supernovae data, whose uncertainty is significantly lower than the one of cosmic chronometers. Hence, at $z = 0$, we do not fit the $H_0$ value, it is imposed as above. 

For the above data and for $\Lambda$CDM with a free $\Omega_{\Lambda 0}$ parameter, we find
\begin{equation}
    \Omega_{\Lambda 0 } = 0.75 \pm 0.02
\end{equation}
at  1$\sigma$ level, which is not identical but close to the WMAP-inferred $\Omega_{\Lambda 0} = 0.72$ \cite{WMAP:2012nax}. For the scale-dependent cosmology, also at 1$\sigma$ level, there is one additional parameter ($\nu$) and we find for the marginalized parameters,
\begin{equation}
     \Omega_{\Lambda 0 } = 0.83^{+0.04}_{-0.05} \quad \mbox{and}  \quad \nu = - 0.6^{+0.9}_{-1.6} \, .
    \label{eq:result}
\end{equation}
 
\begin{figure}[ht!]
\centering
\includegraphics[scale=.3]{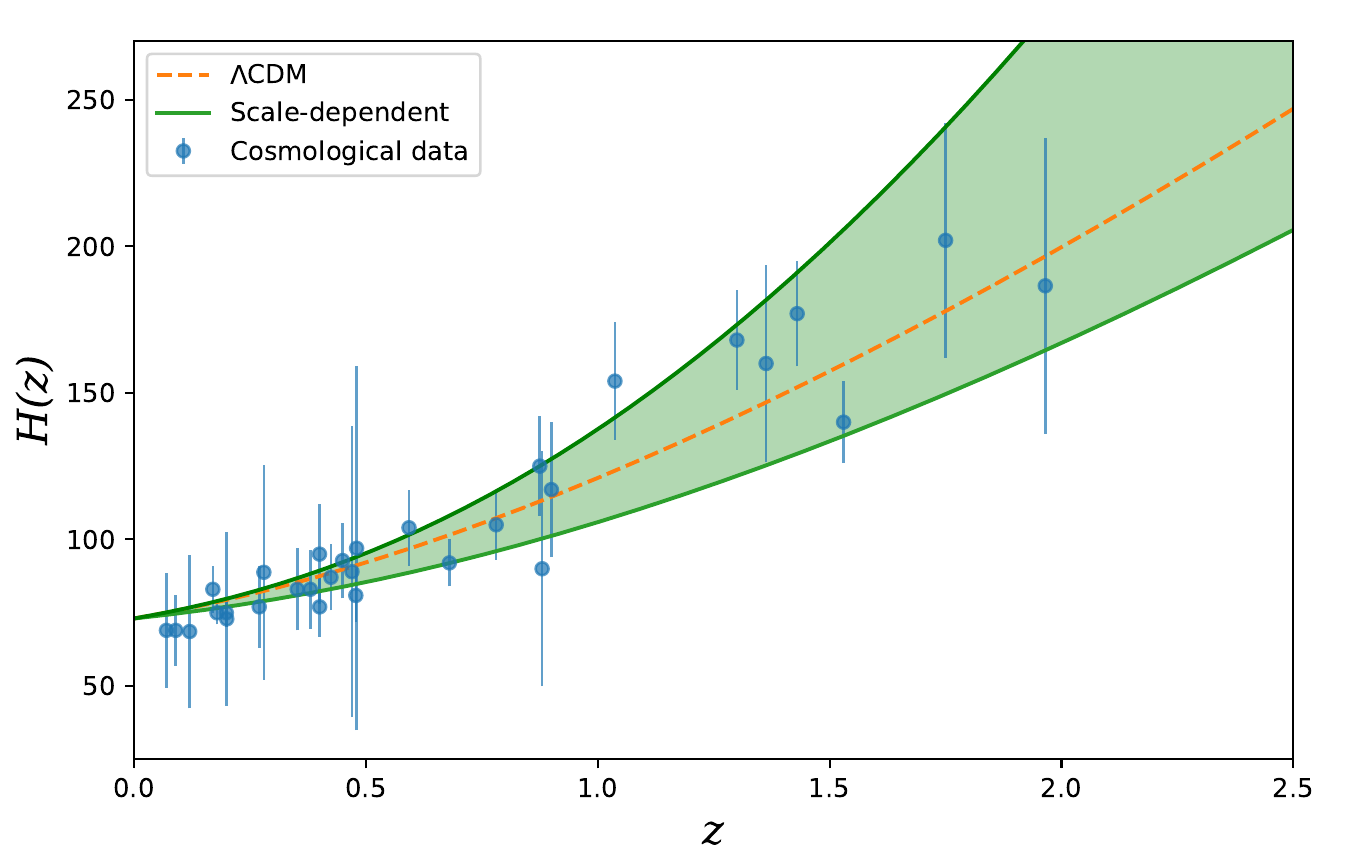}
\caption{Hubble parameter versus redshift. The orange dashed line corresponds to the $\Lambda$CDM best fit ($H_0 = 73.0$ km s$^{-1}$ Mpc$^{-1}$ and $\Omega_{\Lambda0} = 0.75$). For the scale-dependent model (also assuming the same $H_0$ value and considering $\Omega_{\Lambda 0}$ and $\nu$ as free) the green shaded region display its 1$\sigma$ region.}   \label{fig:CsC}
\end{figure}

 The $\chi^2$ minimization results are shown in Table \ref{tab:chi2_result}. It shows that the scale-dependent approach leads to certain fit improvement, favoring negative $\nu$ and larger $\Omega_{\Lambda 0}$ values. The evolution of the RG-corrected Hubble parameter in comparison with the $\Lambda$CDM one is shown in Fig. \ref{fig:CsC}. 

\begin{table}
    \centering
    \caption{Results for $\chi^2$ minimization considering cosmic chronometers data and $H_0 = 73.0$ km s$^{-1}$ Mpc$^{-1}$ \cite{Riess:2021jrx}.}
    \begin{tabular}{c c c}
    \hline
    \hline
         & $\Lambda$CDM & Scale-dependent \\
    \hline
       Best fit          & $\Omega_{\Lambda 0} = 0.75$  & $\Omega_{\Lambda 0} = 0.81$, $\nu = - 1.09$        \\
       $\chi^{2}_{\rm min}$  & 17.12                        & 15.97 \\
    \hline
    \hline
    \end{tabular}
    \label{tab:chi2_result}
\end{table}

The results above favor negative values for $\nu$ that are about the unit. These results are still within the perturbative analysis we do here since the full corrections are still small with respect to the original values. 

Considering the 1$\sigma$ shaded region contrasted with the $\Lambda$CDM result,  the RG correction may alleviate the tension that exists between local estimates of $H_0$ (which yield $h \sim 0.73$)\footnote{Following the standard notation: $H_0 = 100 h$ km s$^{-1}$ Mpc$^{-1}$.} and measurements derived from the cosmic microwave background (with $h\sim 0.68$). Such tension can in principle be alleviated by either considering models whose $H(z)$ evolution at late times becomes flatter; or by significantly enlarging the $H_0$ model uncertainty. A recent review on the Hubble tension can be found in Ref.~\cite{Abdalla:2022yfr}. Nonetheless, we stress that this is at the moment a possible application and a more robust evaluation of this issue requires both theoretical and phenomenological developments. Considering the theoretical part, it is necessary to extend this approach towards high redshift (beyond the weak field regime) and develop the perturbative picture. Cosmological perturbations are not the focus of this work, but we indicate here the direction for further developments.  It is necessary to identify coordinate-independent quantities that correspond to the energy scales $\mu_1$ and $\mu_2$. For instance, Ref.~\cite{CCwave} uses the same scale $\mu_1$ that is used here and sets $\mu_{1}= \nabla_\alpha u^\alpha$, where $u^\alpha$ is the 4-velocity. For $\mu_2$, a possible setting is $\mu_2 =  - \frac{2}{3}u^{\alpha}u^{\beta}\left( R_{\alpha\beta} - \frac{1}{4}g_{\alpha\beta}R  \right)$. With these settings, one can, at least in principle, find the perturbative dynamics. We will address the cosmological perturbations in detail in a future work.

\section{Conclusions and discussions}
\label{secConc}

We extended the framework initially proposed in \cite{UEA-RG}, applying it for the first time to cosmology. Within this framework, the gravitational constant $G$ and the cosmological constant $\Lambda$ exhibit dependency on the energy scale $\mu_1$, as shown in eqs.~\eqref{eq:L} and \eqref{eq:G}. 
The specific running found in ref.~\cite{UEA-RG} satisfies  $G \propto \Lambda^4$, being clearly differently from others (e.g., \cite{CC-Gruni,Bonanno:2001hi, Rodrigues:2015hba}). 
We have shown that, in this context, it is not consistent to simply use the scale $\mu_1 \propto H$ together with energy-momentum conservation. However, introducing a second scale ($\mu_2$) solves the issue, as we have shown here. 

Following ref.~\cite{UEA-RG}, the evolution of $G$, $\Lambda$, and the $R^2$ coupling constant $\xi_2$ should depend solely on $\mu_1$. It is conceivable that higher order $R$-corrections exist, necessitating additional coupling constants. These additional coupling constants can dependent on a secondary scale $\mu_2$. Our analysis introduces $R^3$ and $R^4$ corrections with the coupling constants $\xi_3(\mu_1, \mu_2)$ and $\xi_4(\mu_1, \mu_2)$ respectively, achieving theoretical consistency. As here shown, these additional corrections are not negligible in late-time cosmology.

We have analyzed the late-time background cosmological evolution for arbitrary $\nu$ values, where $\nu = 0$ corresponds to no scale-dependence. 
By analyzing the proposed model with the cosmic chronometers data set \cite{Moresco:2023zys}, we find that the proposed model improves the fit and favors larger $\Omega_{\Lambda 0}$ values and negative $\nu$. These findings hold potential implications for addressing the Hubble tension (e.g.,~\cite{Abdalla:2022yfr}). However, further theoretical investigation is required, particularly during the inflationary period, when quantum corrections may not be negligible. 
We plan to explore these aspects in future work.

\section*{Acknowledgments}
We thank Rodrigo von Marttens for suggesting the use of cosmic chronometers.
NRB acknowledges the support from \textit{Conselho Nacional de Desenvolvimento Cient\'{i}fico e Tecnol\'{o}gico} - CNPq, for the post-doctoral fellowship.
DCR thanks Heidelberg University
for hospitality and support. He also acknowledges partial support from 
\textit{Conselho Nacional de Desenvolvimento Científico e 
Tecnológico} (CNPq-Brazil) and \textit{Fundação de Amparo 
à Pesquisa e Inovação do Espírito Santo} (FAPES-Brazil) (TO 
1020/2022, 976/2022, 1081/2022). I.Sh. acknowledges the partial 
support from CNPq	under the grant 303635/2018-5.

\appendix

\section{Derivation of $\xi_3$ and $\xi_4$} 
\label{AppXi}

Let us describe the procedure for obtaining the functions
\eqref{eq:xi3} and \eqref{eq:xi4} that maintain the conservation of
the energy-momentum tensor in the cosmological context. For the
metric \eqref{eq:metric} and in a comoving frame, the last term in
Eq.~\eqref{eq:div3} is zero and, on top of this,
\ $u^\al u^\be G_{\al\be}=3H^{2}$.
Using the identifications \eqref{eq:ssH}, \eqref{eq:ssdotH} and the function
\eqref{eq:f}, Eq.~\eqref{eq:div3} becomes
\beq
u^\be \,\nabla_{\alpha}T^{\alpha}_{\;\;\beta}
&=&
\dot{H}\bigg[
\frac{d \rho_\Lambda}{d H}
+ \frac{ 3 H^{2}}{8\pi G^2} \frac{d G}{d H}
- \frac{1}{16\pi}\Big(\frac{d\xi}{dH}R^{2}
+ \frac{\partial \xi_3}{\partial H}R^{3}
+ \frac{\partial \xi_4}{\partial H}R^{4}\Big) \bigg]
\nn
\\
&&
-\,\,
\frac{\ddot{H}}{16\pi}\bigg( \frac{\partial\xi_3}{\partial\dot{H}}R^{3}
+  \frac{\partial\xi_4}{\partial\dot{H}} R^{4}\Big).
\eeq
For $u^{\beta}\nabla_{\alpha}T^{\alpha}_{\;\;\beta} =0$, the
coefficients of $\dot{H}$ and $\ddot{H}$ must simultaneously be
zero. Therefore, using \eqref{eq:L}, \eqref{eq:G} and \eqref{eq:xi}
and denoting $x=H$ and $y=\dot{H}$ to simplify the notations, the
coefficient of $\dot{H}$ is
\begin{align}
\frac{\alpha}{x} + \beta x + \frac{\sigma}{x}R^2
- \frac{\partial \xi_3}{\partial x}R^3
- \frac{\partial \xi_4}{\partial x}R^{4}
\,=\, 0.
\label{eq:dif1}
\end{align}
The coefficient of $\ddot{H}$ is
\begin{align}
\frac{\partial\xi_3}{\partial y}R^3
+ \frac{\partial \xi_4}{\partial y}R^{4}
\,=\, 0,
\label{eq:dif2}
\end{align}
where
\begin{align}
\label{eq:const}
\alpha = \frac{6\Lambda_0 \nu}{5G_0},
\quad
\beta = -\frac{24\nu}{5G_0},
\quad
\sigma = - 16\pi \xi_0 \kappa ,
\quad
R = 6 (2x^{2}+y).
\end{align}
The system of partial differential equations presented above can be
solved analytically. It is convenient to find the equations involving
only $\xi_3$, and only $\xi_4$.  In the second case, this can be
done as follows. By differentiating \eqref{eq:dif1} with respect
to $y$, we get
\begin{align}
\frac{2\sigma}{x}R\frac{\partial R}{\partial y}
- R^3\frac{\partial^{2} \xi_3}{\partial y \partial x}
- 3R^{2}\frac{\partial\xi_3}{\partial x}\frac{\partial R}{\partial y}
- \frac{\partial^{2}\xi_4}{\partial y \partial x}R^4
- 4R^{3}\frac{\partial \xi_4}{\partial x}\frac{\partial R}{\partial y}
= 0,
\label{eq:dif3}
\end{align}
Differentiating the Eq.~\eqref{eq:dif2} with respect to $x$ and using
this same equation to eliminate $\pa \xi_3/\pa y$, results in
\begin{align}
-\frac{\partial^{2} \xi_3}{\partial x \partial y}R^3
\,=\,
R^{3}\frac{\partial\xi_4}{\partial y}\frac{\partial R}{\partial x}
+ R^{4}\frac{\partial^{2} \xi_4}{\partial x \partial y} .
\label{eq:dif2x}
\end{align}
On the other hand, Eq.~\eqref{eq:dif1} can be rewritten as
\begin{align}
-\frac{\partial \xi_3}{\partial x}R^2
\,=\, \frac{\partial\xi_4}{\partial x}R^{3}
- \frac{\sigma}{x}R - \beta \frac{x}{R}-\frac{\alpha}{x R}.
\label{eq:dif12}
\end{align}
Using \eqref{eq:dif2x} and \eqref{eq:dif12} in
\eqref{eq:dif3} with the conditions
\begin{align}
\frac{\partial ^2 \xi_3}{\partial x\partial y}
\,=\,
\frac{\partial ^2 \xi_3}{\partial y\partial x},
\qquad
\frac{\partial ^2 \xi_4}{\partial x\partial y}
\,=\,
 \frac{\partial ^2 \xi_4}{\partial y\partial x},
\end{align}
we get
\begin{align}
R^3 \left( \frac{\partial \xi_4}{\partial y}\frac{\partial R}{\partial x}
- \frac{\partial\xi_4}{\partial x}\frac{\partial R}{\partial y} \right)
- \left( \frac{\sigma R}{x} + \frac{3\beta x}{R}
+ \frac{3\alpha}{xR} \right)\frac{\partial R}{\partial y} \,=\, 0.
\end{align}
Solving this equation by requiring $\xi_4$ to be a constant,
$\xi_{40}$, corresponding to $x=x_0$ (i.e., to the fiducial scale
$\mu_0$), we arrive at
\begin{align}
\label{eq:xi4xy}
\xi_4(x,y) \,=\,
\xi_{40} - \frac{\beta (x^{2}-x_0^{2})}{864(2x^{2}+y)^2}
- \frac{\alpha + 12(2x^2+y)\sigma}{432(2x^2+y)^4}
\,\ln \Big(\frac{x}{x_0}\Big).
\end{align}
Taking this function in \eqref{eq:dif2} and integrating over $y$ results in
\begin{align}\label{eq:xi3xy}
\xi_3(x,y)
\,=\,
\frac{\beta(x^{2} - x_0^2)}{108(2x^{2}+y)^{3}}
+\frac{\alpha +18\sigma(2x^2+y)^{2}}{54(2x^{2}+y)^3}
\,\ln \Big(\frac{x}{x_0}\Big) + \xi_{30}(x),
\end{align}
where $\xi_{30}(x)$ is an arbitrary function.
Inserting this result into \eqref{eq:dif1} gives the condition
$d\xi_{30}/dx = 0$, then $\xi_{30}$ must be constant. The
functions \eqref{eq:xi3xy} and \eqref{eq:xi4xy} with the
constants $\alpha$, $\beta$ and $\sigma$ in \eqref{eq:const}
result in \eqref{eq:xi4} and \eqref{eq:xi3}, respectively.

\end{document}